\documentclass[useAMS,usenatbib]{mn2e}
\usepackage{setspace}

\usepackage{caption}
\usepackage[english]{babel}
\usepackage{graphicx}
\usepackage{subfigure}

\usepackage{rotating}
\usepackage{stfloats}
\usepackage{txfonts} 
\usepackage{textcomp}	
\usepackage{natbib}
\usepackage{amssymb}
\usepackage{graphicx}
\usepackage{tabularx}
\usepackage[english]{babel}
\usepackage[T1]{fontenc}
\usepackage{multirow}

\usepackage{threeparttable}

\usepackage{graphics}
\usepackage{graphpap}
\usepackage{rotating}
\usepackage{color}

\def\HI{H{\,\small I}}

\newcommand{\kms}{$\,$km$\,$s$^{-1}$}

\newcommand{\msun}{{${\rm M}_\odot$}}

\def\HI{H{\,\small I}}

\def\emph#1{{\sl #1}}
\newcommand{\ltsima} {$\; \buildrel < \over \sim \;$}
\newcommand{\gtsima} {$\; \buildrel > \over \sim \;$}
\newcommand{\lta} {\lower.5ex\hbox{\ltsima}}
\newcommand{\gta} {\lower.5ex\hbox{\gtsima}}

\newcommand{\atlas}{{ATLAS$^{\rm 3D}$}}

\title[A multiwavelength survey of HI-excess galaxies]{A multiwavelength survey of HI-excess galaxies with surprisingly inefficient star formation}
\author[K. Ger\'{e}b et al.]
{K. Ger\'{e}b$^1$\thanks{E-mail:
kgereb@swin.edu.au},
S. Janowiecki$^{2}$,
B. Catinella$^{2}$,
L. Cortese$^{2}$,
V. Kilborn$^{1}$
\\
$^1$Centre for Astrophysics and Supercomputing, Swinburne University of Technology, Hawthorn, VIC 3122, Australia\\
$^2$ICRAR, M468, The University of Western Australia, 35 Stirling Highway, Crawley, Western Australia, 6009, Australia}

\begin{document}


\pagerange{\pageref{firstpage}--\pageref{lastpage}} \pubyear{2017}

\maketitle

\label{firstpage}

\begin{abstract}
We present the results of a multiwavelength survey of \HI-excess galaxies, an intriguing population with large \HI\ reservoirs associated with little current star formation. 
These galaxies have stellar masses $M_{\star} >10^{10}$ M$_{\odot}$, and were identified as outliers in the gas fraction vs. NUV$-r$ color and stellar mass surface density scaling relations based on the GALEX Arecibo SDSS Survey (GASS). We obtained \HI\ interferometry with the GMRT, Keck optical long-slit spectroscopy and deep optical imaging (where available) for four galaxies.
Our analysis reveals multiple possible reasons for the \HI\ excess in these systems. One galaxy, AGC 10111, shows an \HI\ disk that is counter-rotating with respect to the stellar bulge, a clear indication of external origin of the gas. Another galaxy appears to host a Malin 1-type disk, where a large specific angular momentum has to be invoked to explain the extreme $M_{\rm HI}$/$M_{\star}$ ratio of 166$\%$. 
The other two galaxies have early-type morphology with very high gas fractions. The lack of merger signatures (unsettled gas, stellar shells and streams) in these systems suggests that these gas-rich disks have been built several Gyr-s ago, but it remains unclear how the gas reservoirs were assembled. Numerical simulations of large cosmological volumes are needed to gain insight into the formation of these rare and interesting systems.

\end{abstract}



\section{Introduction}\label{Sec:Intro}

Numerical simulations predict that cosmological accretion provides fresh supplies of gas that form stars in galaxies at high redshift \citep{Binney2004, Keres2005, Dekel2009}. 
In the nearby Universe, star-forming galaxies can accrete gas from their halos via the galactic fountain mechanism \citep{Fraternali2002}. Gas-rich mergers and infall from the intergalactic medium have been suggested as sources of fresh gas supplies \citep{Sancisi2008}, and more recent studies highlight the importance of galactic winds as well \citep{Angles2017}. 
Once the gas reaches the inner regions of a galaxy (within a few effective radii) and settles into a disk, the signatures of gas accretion (e.g., scattered clouds, streams, tails) will disappear soon after completing 1-2 orbits (equal to a few Gyr for massive galaxies, \citealt{vanWoerden1983, Serra2006}), and the gas will be converted into stars. Given the interplay between gas and star formation, a well known correlation \citep{Schmidt1959, Kennicutt1989, Kennicutt1998} exists between gas (either \HI+H$_2$ or H$_2$) and star formation rate (SFR) densities. This is commonly observed in star-forming disks where the density of \HI\ is high enough to be converted into molecular gas. However, this relation breaks down at low gas densities, where star-forming processes are highly inefficient. This is particularly relevant at galaxy outskirts \citep{Bush2008, Leroy2008, Bigiel2010, Yildiz2015, Yildiz2017}, where the gravitational potential is very shallow and high angular momentum may prevent the collapse of gas \citep{Kim2013, Obreschkow2016, Lutz2017}. Thus, if the gas is unable to reach the central regions, it will remain preserved in atomic form at the galaxy outskirts, resulting in a disconnection between global \HI\ content and SFR.

Previous studies have looked into the role of specific angular momentum in the accumulation of massive \HI\ reservoirs in unusually gas-rich galaxies. A classic example of this is Malin 1 \citep{Pickering1997,Lelli2010}. The specific angular momentum of this galaxy is $\sim$20 times larger than that of the Milky Way \citep{Boissier2016}, spreading its $10^{10.82}$ \msun\ of \HI\ gas into an extended, 130 kpc radius disk.  
There are indications that angular momentum plays an important role in galaxies that are \HI-rich for their stellar content (mass, luminosity, see \citealt{Huang2014, Lemonias2014, Lutz2017}). These studies have all suggested high angular momentum (or in some cases the spin parameter) as a possible mechanism to explain the observed properties (unusually extended disks, low gas surface densities) of such \HI-rich galaxies.  
But are all \HI-rich galaxies the same?

In our previous work (\citealt{Gereb2016}, Paper I from now on), we identified a rare population of so-called \HI-excess systems, which might be good candidates for galaxies that have recently accreted gas. These are characterised by high gas fractions ($M_{\rm HI}/M_{\star}$) paired with low specific SFRs (estimated based on the combination of NUV$-r$ color and stellar mass surface density, $\mu_{\star}$) in the GASS sample \citep{Catinella2010, Catinella2013}. The prototype galaxy of this class, GASS 3505, was analysed in detail in Paper I. The contrast between its 10$^{9.9}$ \msun\ of \HI\ mass and low level of star formation (0.1 $M_{\odot}$ yr$^{-1}$) pushes GASS 3505 into the \HI-excess regime. We carried out numerical simulations and showed that a GASS 3505-type system can be formed in a merger between a very gas-rich dwarf and a bulge-dominated galaxy. However, GASS 3505 is only one example, and a larger sample needs to be explored in order to understand the \HI-excess population as a whole.

We are carrying out an observing campaign with the Giant Metrewave Radio Telescope (GMRT) to image the gas distribution of \HI-excess galaxies. In this paper we present the GMRT observations of four \HI-excess galaxies along with Very Large Telescope (VLT) and Canada France Hawaii Telescope (CFHT) deep optical imaging and Keck spectroscopy from dedicated follow-up observations. In addition to the above, Sloan Digital Sky Survey (SDSS; \citealt{York}) photometry and spectroscopy, furthermore Galaxy Evolution Explorer (GALEX; \citealt{Martin2005}) imaging are also available by selection. 
These multiwavelength data allow us to explore the \HI-excess in relation to star formation, structure, and gas accretion signatures in the sample, and bring us a step closer to understanding the evolutionary processes in these galaxies. 
In this paper the standard cosmological model is used with parameters $\Omega_{\rm M}$ = 0.3, $\Omega_{\Lambda}$ = 0.7 and $H_0$ = 70 km s$^{-1}$ Mpc$^{-1}$.

\begin{figure}
\includegraphics[width=.5\textwidth]{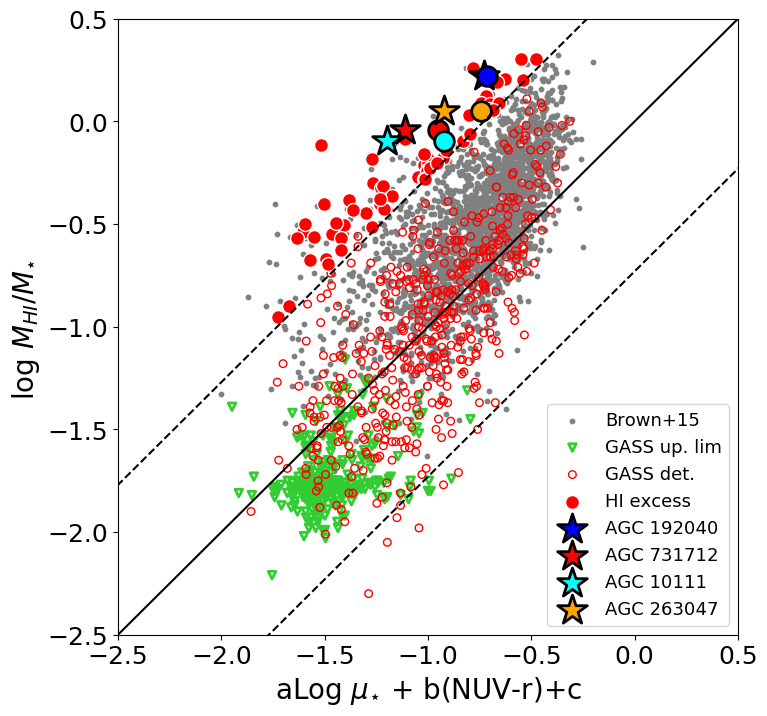}
\begin{center}\caption{Gas fraction plane for ALFALFA and GASS galaxies with $M_\star > 10^{10}$ M$_{\odot}$.
Red, empty circles indicate the GASS sample, grey points represent ALFALFA galaxies, and filled red circles show the \HI-excess sample. Green triangles are the GASS upper limits. Star symbols mark the four \HI-excess galaxies studied in this paper, and coloured circles represent the revised NUV$-r$ measurements for these, as discussed in Sec. \ref{sec:color_mustar}. The solid line represents the 1:1 relation, and the dashed lines the 2.5$\sigma$ deviation from it. The values of the coefficients are a = -0.240, b = -0.250, c = 2.083.
}\label{fig:GFplane}
\end{center}
\end{figure}

\begin{figure*}
\begin{picture}(600,600)(0,0)
\put(-25,460){\includegraphics[width=.2\textwidth]{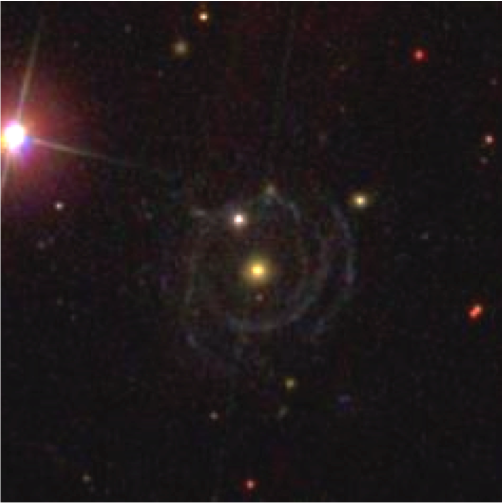}}
\put(84,443){\includegraphics[width=.91\textwidth]{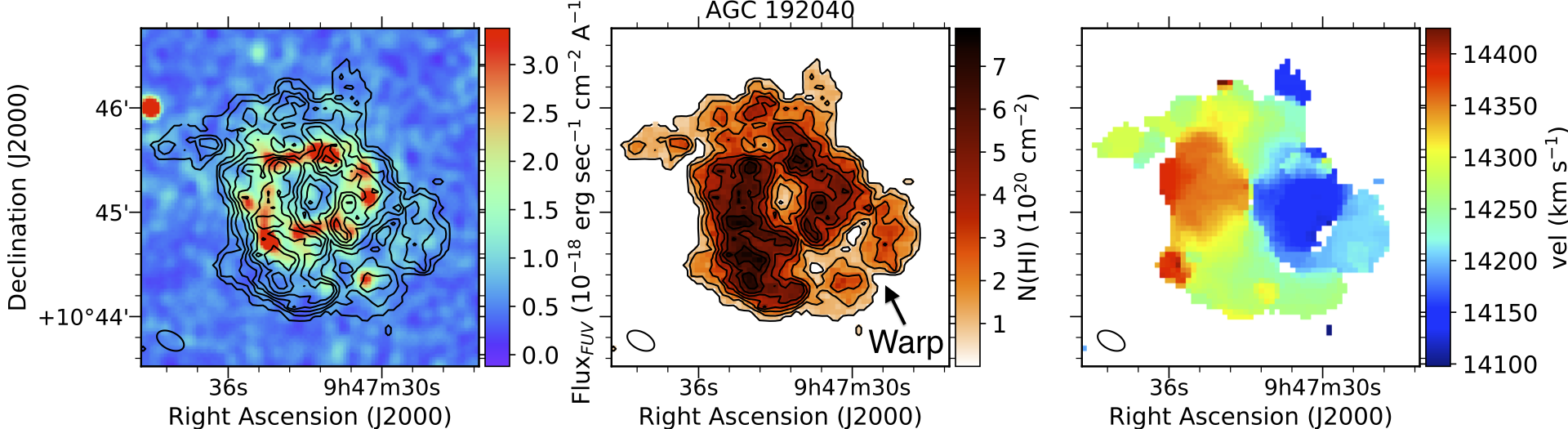}}
\put(-25,321){\includegraphics[width=.2\textwidth]{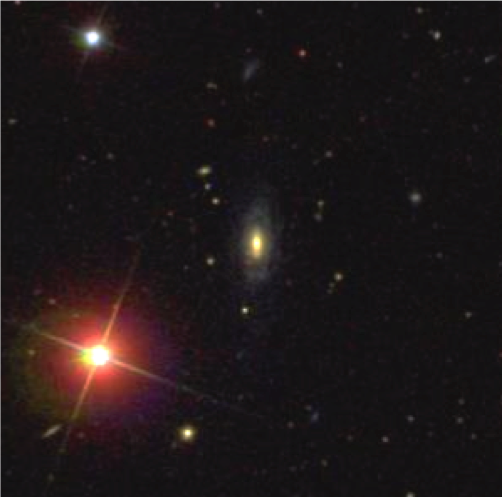}}
\put(84,303){\includegraphics[width=.91\textwidth]{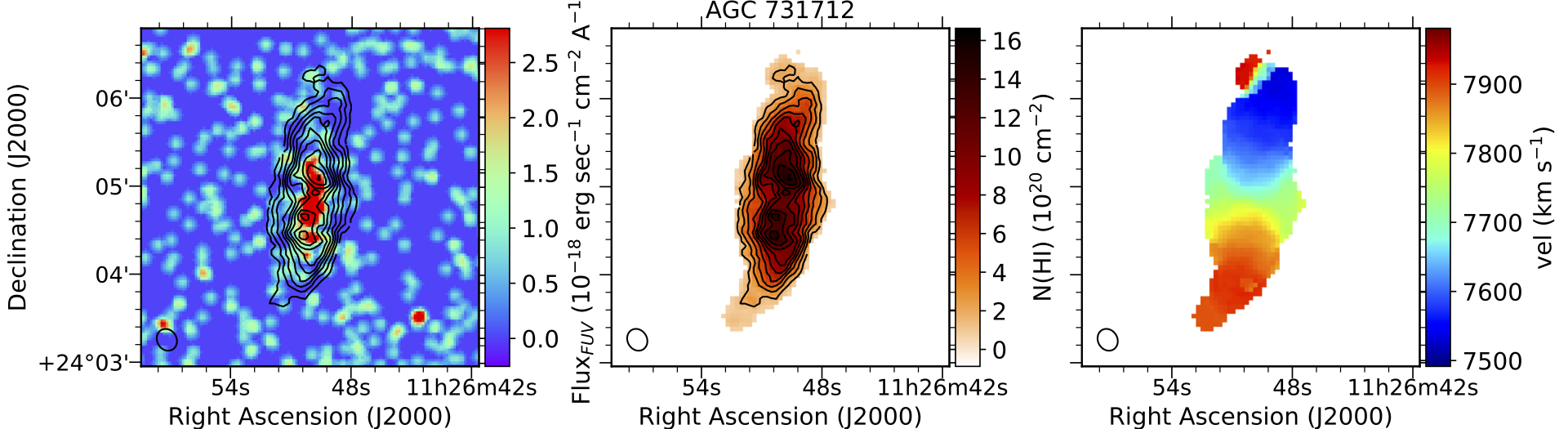}}
\put(-25,180){\includegraphics[width=.2\textwidth]{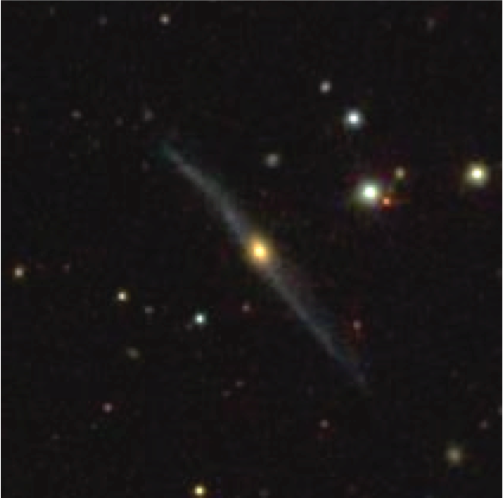}}
\put(84,163){\includegraphics[width=.915\textwidth]{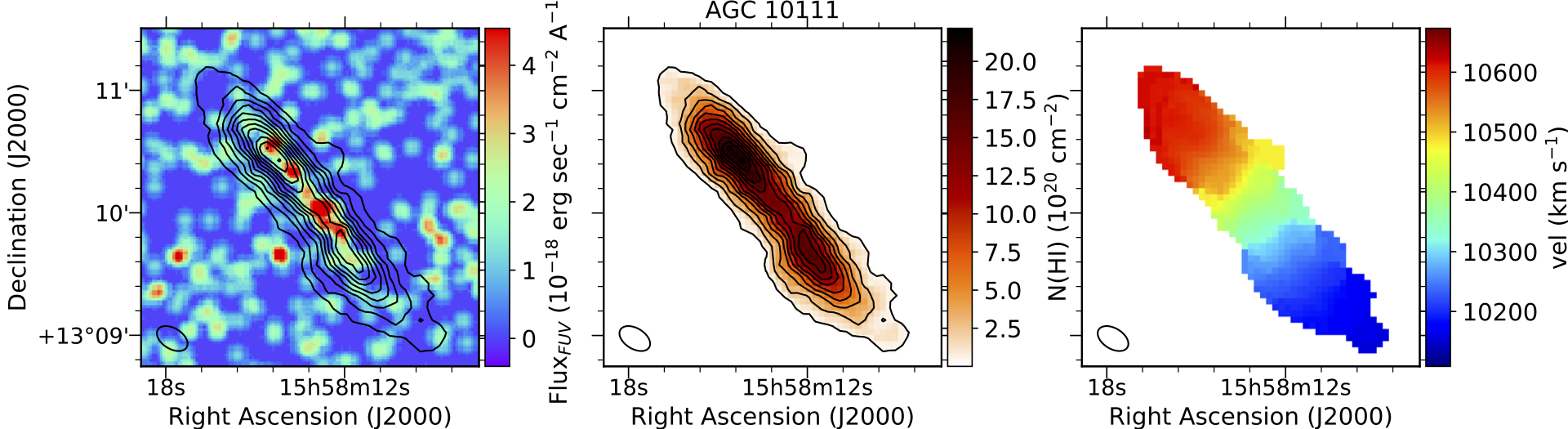}}
\put(-25,40){\includegraphics[width=.2\textwidth]{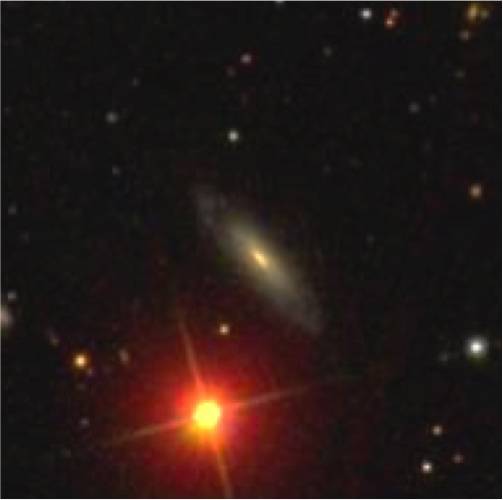}}
\put(84,23){\includegraphics[width=.91\textwidth]{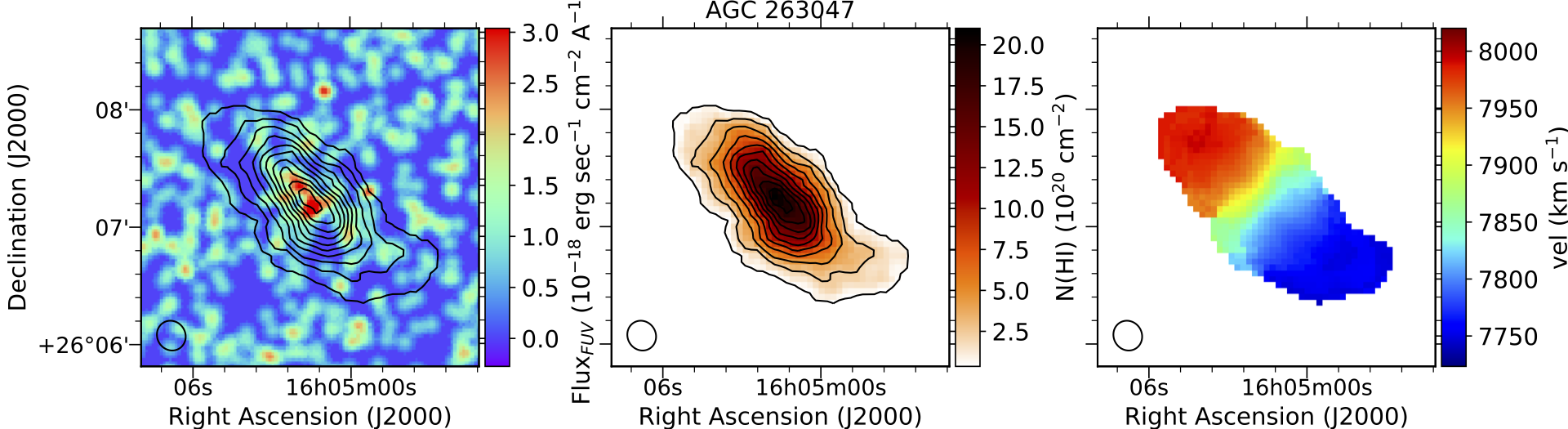}}
\end{picture}
\caption{Multiwavelength observations of the \HI-excess galaxies. From left to right: SDSS optical image, FUV image overlayed with \HI\ contours, \HI\ column density, \HI\ velocity field. The SDSS images are extrated on a scale similar to the FUV and \HI\ maps. The FUV images are smoothed by a Gaussian kernel ($r=4$ pix) for illustration purposes. The column density contours range between $\sim3\sigma$ and maximum intensity, where the steps are defined depending on the dynamic range. The contours are set to be (0.35 - 9)$\times$10$^{20}$ cm$^{-2}$ with steps of 1.05$\times$10$^{20}$ cm$^{-2}$ for AGC 192040, (0.3 - 17)$\times$10$^{20}$ cm$^{-2}$ with steps of 1.5$\times$10$^{20}$ cm$^{-2}$ for AGC 731712, (0.35 - 22)$\times$10$^{20}$ cm$^{-2}$ with steps of 2.1$\times$10$^{20}$ cm$^{-2}$ for AGC 10111 and AGC 263047. The major axis of the radio beam (black ellipses) in the images is 15$\pm$1 arcsec. 
} \label{fig:Collection}
\end{figure*}

\vspace{-0.3cm}

\section{Sample Selection}\label{Sec:SampleSelection}

Our sample is selected from the GALEX Arecibo SDSS Survey (GASS, \citealt{Catinella2010, Catinella2013}) based on the gas fraction plane, a relation between the measured \HI\ gas fraction and that predicted from the combination of NUV$-r$ color (a proxy for specific SFR) and stellar mass surface density ($\mu_{\star}$, indicator of morphological type), as shown in Fig. \ref{fig:GFplane}. 
From the GASS sample we selected all the galaxies with a gas fraction in excess of $2.5{\sigma}$ from the average relation in Fig. \ref{fig:GFplane}, and examined the SDSS images of each of them to exclude those affected by \HI\ confusion within the 3.5 arcmin beam of the Arecibo telescope (shown as grey points above the upper dashed line in Fig. \ref{fig:GFplane}). This yields 36 galaxies. In order to maximise the number of \HI-excess galaxies, which are rare but very \HI-rich for their star formation and structure, here we combine GASS with the sample of \cite{Brown2015}. The latter sample is extracted from SDSS and cross-matched with the $\alpha.70$ source catalog from ALFALFA \citep{Giovanelli2005, Haynes2011}, and contains $\sim11230$ galaxies in the redshift and stellar mass ranges of GASS ($0.025 < z < 0.05$, $M_{\star} > 10^{10}$ M$_{\odot}$). From this sample we selected \HI-excess systems not already included in GASS and excluded confused ones, obtaining 28 additional galaxies. Thus, our \HI-excess sample includes a total of 64 galaxies.

For both samples the stellar masses are taken from the MPA/JHU value added catalogs, based on SDSS DR7\footnote{http://www.mpa-garching.mpg.de/SDSS/DR7/}. The stellar mass surface density is defined as $\mu_{\star}$ = $M_\star$ /(2$\pi$ $r_{50,z}^2$), where $r_{50,z}$ is the radius containing 50$\%$ of the Petrosian flux in the SDSS $z$-band in kpc. We note however that the GALEX UV photometry for GASS was reprocessed as in \cite{Wang2010}, whereas the \cite{Brown2015} sample is taken from the GALEX Unique Source Catalogues\footnote{http://archive.stsci.edu/prepds/gcat/} (GCAT, \citealt{Seibert2012}) and the \cite{Bianchi2014} catalog of Unique GALEX Sources\footnote{http://archive.stsci.edu/prepds/bcscat/} (BCS).
For more details on the parent sample descriptions see \cite{Catinella2010} and \cite{Brown2015}.

The 64 \HI-excess galaxies above the $2.5\sigma$ line are marked as red filled circles in Fig. \ref{fig:GFplane}. In this paper we focus on the detailed analysis of four targets marked with coloured star symbols.
These are selected to show signatures of recent star formation, i.e., low surface brightness arms, disks in the SDSS images for optical spectroscopic observation purposes, see left panel of Fig. \ref{fig:Collection}. 
The galaxies are identified by their Arecibo General Catalog (AGC, maintained by M.P. Haynes and R. Giovanelli at Cornell University) number, and their main properties are summarized in Table \ref{table:summary}. 

In what follows, we take advantage of our deep optical, GMRT and Keck follow-up observations to shed light on the physical processes that are responsible for the presence of excess \HI\ gas in these galaxies.

\begin{table*}
  \caption{Basic properties of the \HI-excess galaxies. Col.1 name used in this paper; Col.2 right ascension (J2000); Col.3 declination (J2000); Col.4 SDSS redshift; Col.5 stellar mass surface density measured from SDSS; Col.6 NUV-r color based on GALEX and SDSS measurements; Col.7; Stellar mass ; Col.8 Arecibo \HI\ mass; Col.9 global SFR measured within the FUV radius, see Sec. \ref{Sec:UVprop} for details.}\label{table:summary}
  
   \begin{center}
     \begin{tabular}{c c c c c c c c c}		
\hline									     
Name &  RA   & Dec   & z & log$\mu _{\star}$      & NUV $-$ $r$ & log$M_{\star}$ & log$M_{\rm \HI}$  & SFR \\ 
     & (deg) & (deg) &   & (M$_{\odot}$ kpc$^{-2}$) & (mag)       &   (\msun)    &  (\msun)          &  (\msun yr$^{-1}$) \\
\hline	
AGC 192040  & 09:47:32.79   & +10:45:08.72  & 0.0475  & 9.4  & 2.23  & 10.54  & 10.76  & 1.33  \\ 
AGC 731712  & 11:26:50.06   & +24:04:52.89  & 0.0257  & 9.26 & 3.89  & 10.16  & 10.12  & 0.24 \\
AGC 10111   & 15:58:13.16   & +13:10:07.80  & 0.0346  & 9.38 & 4.12  & 10.32  & 10.23  & 0.14 \\
AGC 263047  & 16:05:01.53   & +26:07:15.03  & 0.0263  & 8.44 & 3.92  & 10.03  & 10.08  & 0.05 \\
\hline
\end{tabular}
\end{center}
\end{table*}


\vspace{-0.3cm}

\begin{figure*}
\begin{picture}(300,250)(0,0)
\put(-110,40){\includegraphics[width=.35\textwidth]{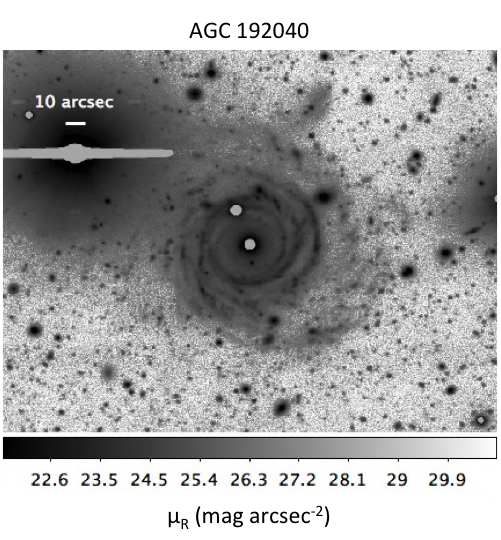}}
\put(65,39.5){\includegraphics[width=.35\textwidth]{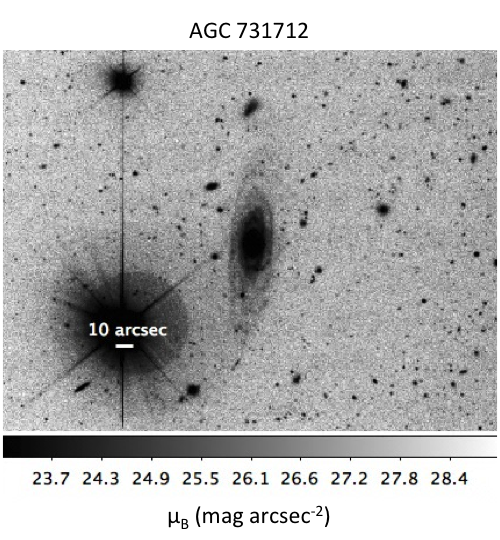}}
\put(240,39.5){\includegraphics[width=.349\textwidth]{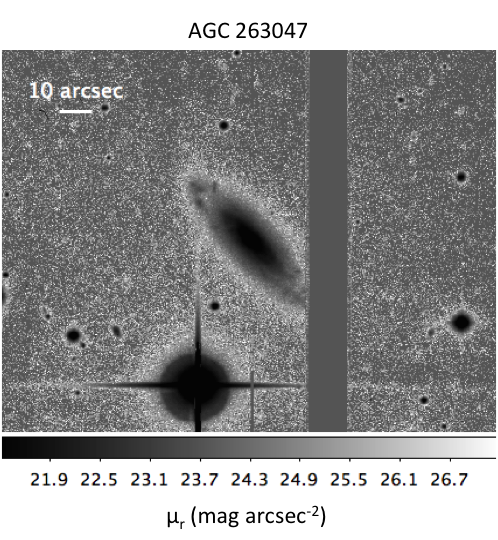}}
\end{picture}
\caption{Deep optical images of AGC 192040 (VLT, left), AGC 731712 (CFHT, middle), and AGC 263047 (CFHT, right). The colourbars show the surface brightness ($\mu$) in the corresponding bands.}\label{fig:Collection_Optical}
\end{figure*}

\section{Observations and data reduction}\label{Sec:DataReduction}

\subsection{Deep optical imaging}

We obtained deep optical images for three of the galaxies in this sample. Very Large Telsecope - FOcal Reducer/low dispersion Spectrograph 2 (VLT/FORS2) observations of AGC 192040 were undertaken in 2011-2012, and include 15x240s exposures in the R filter. These observations were reduced with the standard European Southern Observatory (ESO) pipeline in the Reflex environment \citep{Freudling2013}. Archival images for AGC 731712 and AGC 263047 were obtained from the CFHT archive\footnote{$http://www.cadc-ccda.hia-iha.nrc-cnrc.gc.ca/en/search/$}. AGC 731712 was observed in B and V filters for 840s and 480s respectively, by the CFH12K Mosaic camera in 2001; AGC 263047 was observed in the r filter for 3x240s with the MegaPrime camera in 2008. Standard Level 2 calibrated data were obtained through the Canadian Astronomy Data Centre (CADC). An additional dark sky flat field correction was applied to observations of all three galaxies and photometric calibrations were made with foreground stars measured by SDSS DR9 \citep{Ahn2012}, following the methodology of \cite{Janowiecki2015}. The resulting three images are shown in Fig. \ref{fig:Collection_Optical}. We note that no deep images are available for AGC 10111.

\begin{figure*}
\begin{picture}(500,120)(0,0)
\put(5,0){\includegraphics[width=.235\textwidth]{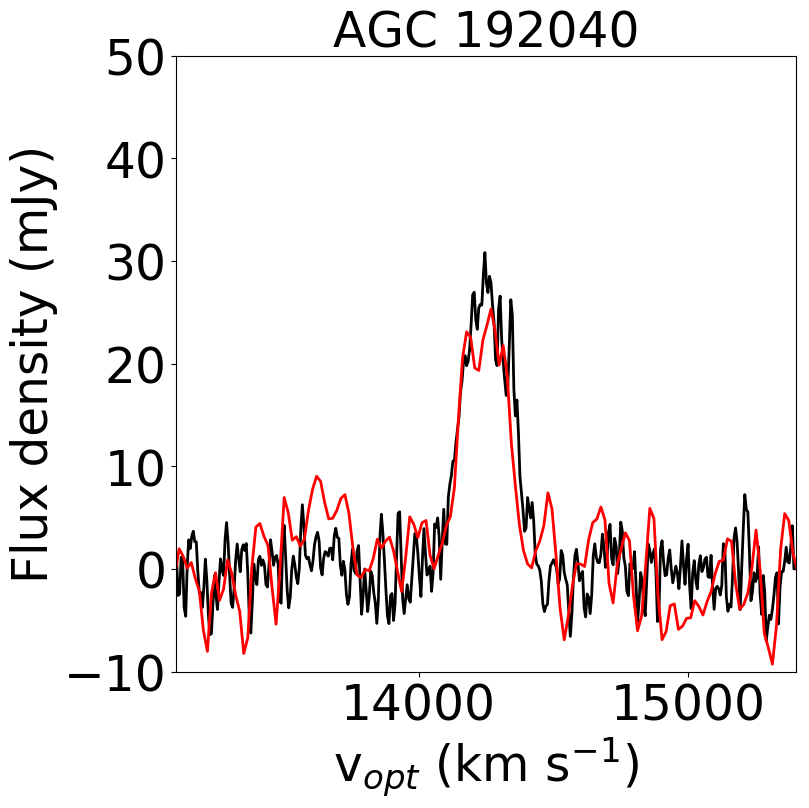}}
\put(131,0){\includegraphics[width=.241\textwidth]{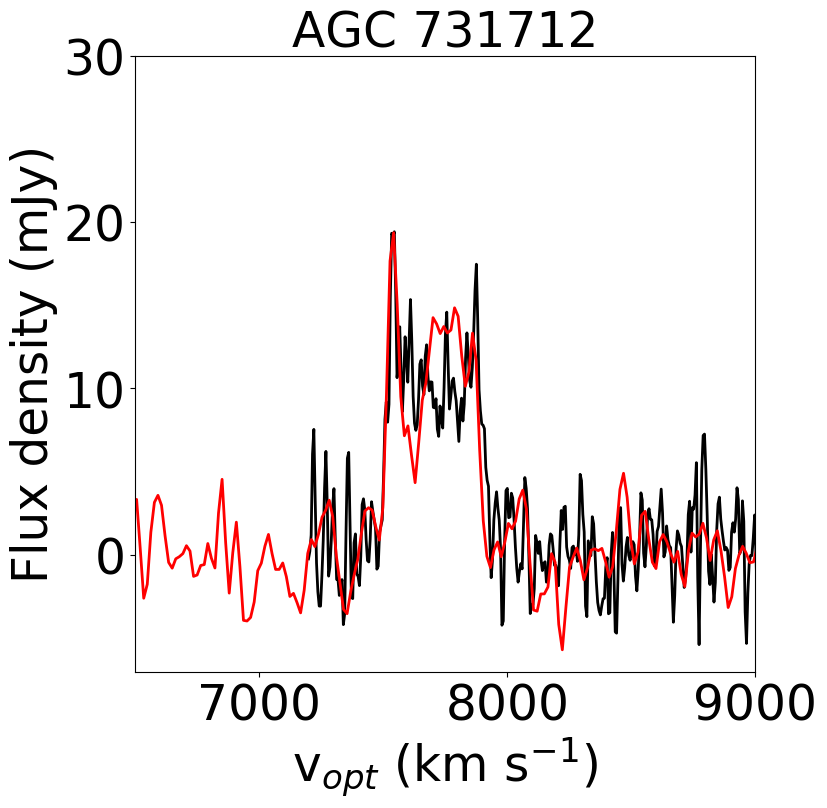}}
\put(253,-1){\includegraphics[width=.228\textwidth]{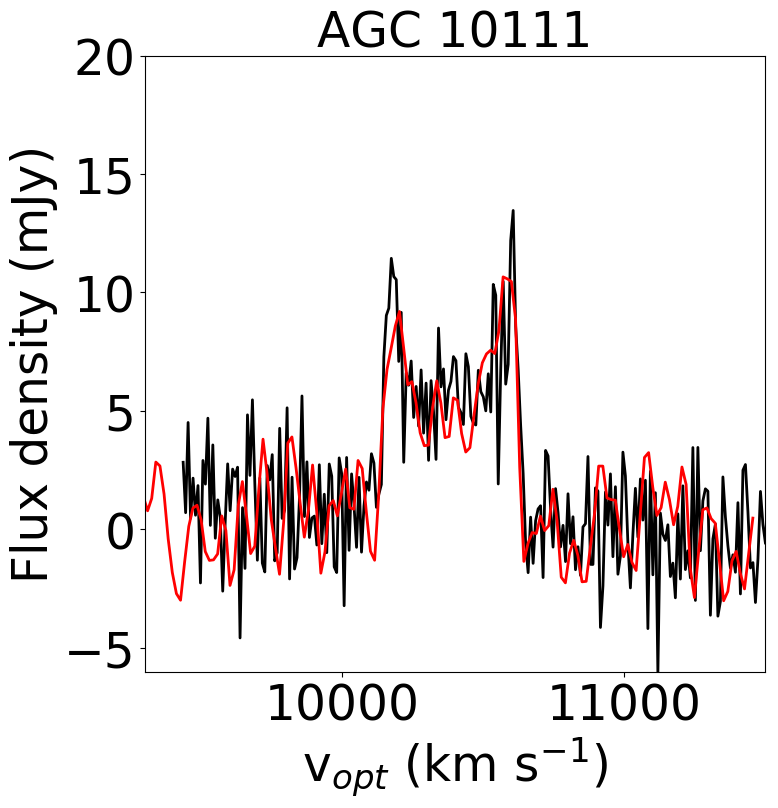}}
\put(374,-2){\includegraphics[width=.226\textwidth]{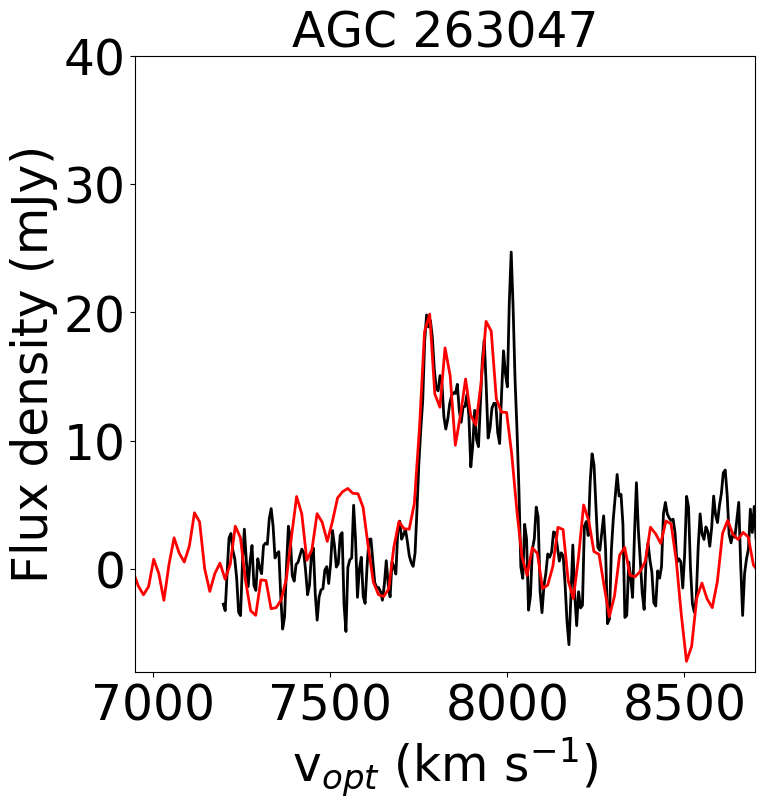}}
\end{picture}
\begin{center}\caption{Comparison between the Arecibo (black line) and GMRT (red line) \HI\ spectra. 
The single-dish spectrum of AGC 192040 is from GASS, the other three galaxies are extracted from ALFALFA.
Galaxy identifiers are noted at the top of each spectrum.}\label{fig:Spectra}
\end{center}
\end{figure*}

\begin{figure*}
\begin{picture}(500,320)(0,0)
\put(30,180){\includegraphics[width=.44\textwidth]{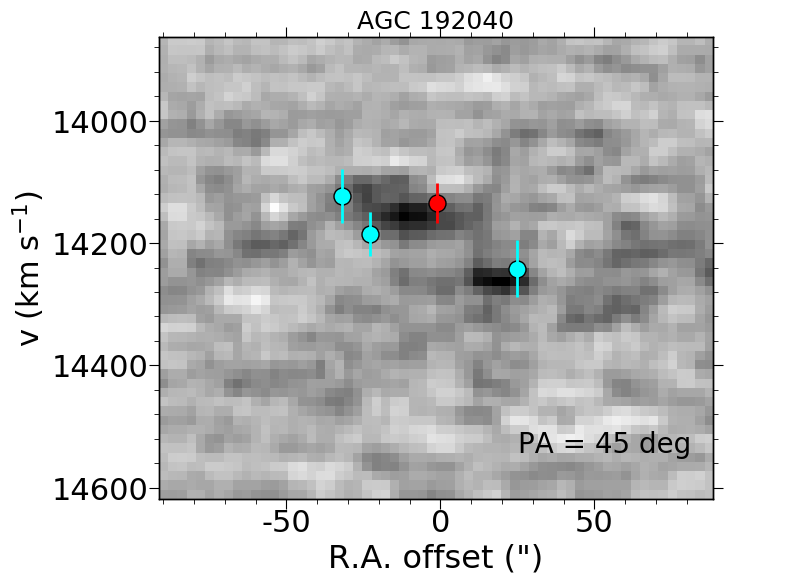}}
\put(260,182){\includegraphics[width=.43\textwidth]{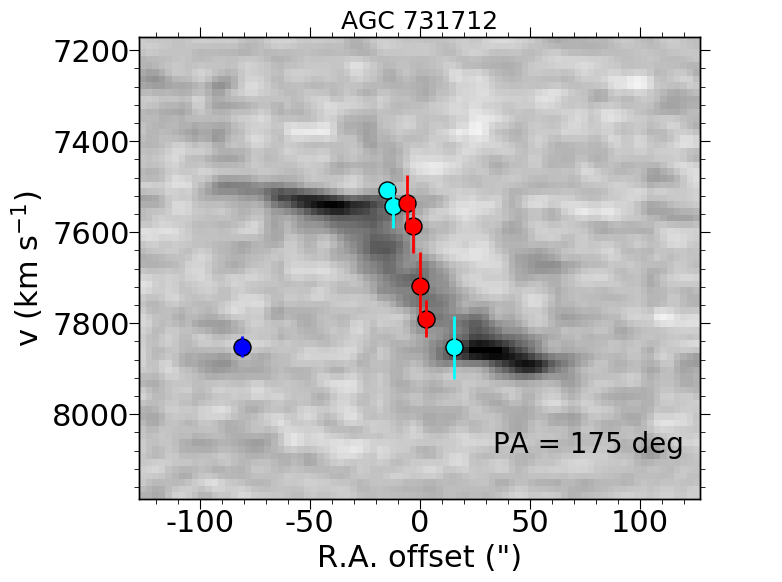}}
\put(30,5){\includegraphics[width=.45\textwidth]{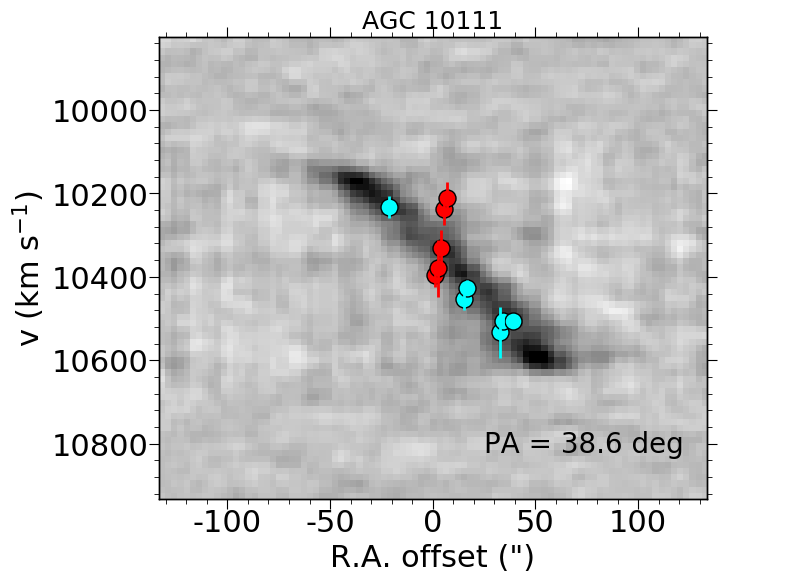}}
\put(260,3){\includegraphics[width=.44\textwidth]{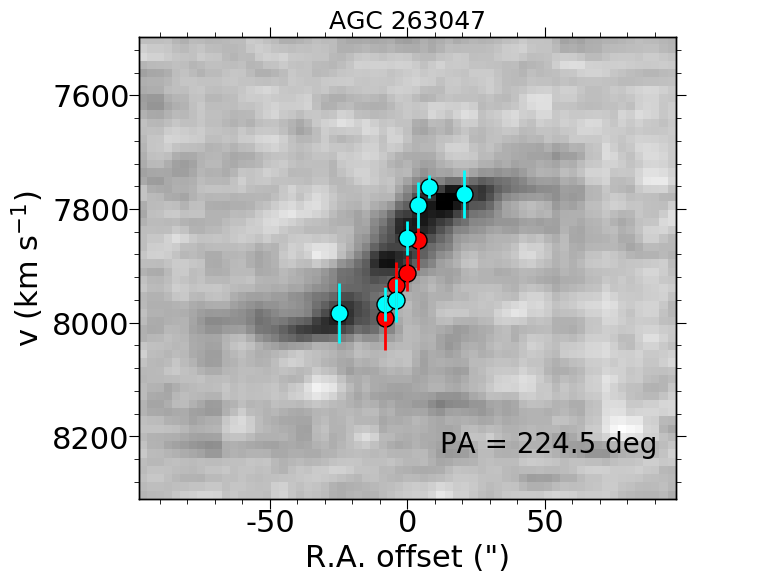}}
\end{picture}
\caption{\HI\ position-velocity diagrams extracted along the position angles of the Keck observations. The \HI\ is shown by greyscales. Measurements based on the H$\beta$ (4861.3$\AA$), H$\gamma$ (4340.5$\AA$), and OIII (4958.9$\AA$, 5006.8$\AA$) emission lines are shown as cyan points. The NaI (5892.5$\AA$) and E-band (5269.0$\AA$) absorption line measurements are marked in red. In the diagram of AGC 731712 (top right) the dark blue point marks the Keck detection of a dwarf galaxy. This galaxy appears along PA = 182 deg and therefore the underlying \HI\ signal is not visible in this plot. The dwarf galaxy is marked with red contours in the north-eastern part of AGC 731712 in Fig. \ref{fig:All_env} (middle panel), and it is also visible in the same position in Fig. \ref{fig:Collection}. } \label{fig:HI_pv}
\end{figure*}

\subsection{Interferometric HI observations}

GMRT \HI\ observations were carried out between Aug 2015 - Dec 2016. Each galaxy was observed for 25 hours, except for AGC 263047 where only 7 hours of observations are available. The 16 MHz bandwidth was covered by 512 frequency channels corresponding to a velocity resolution of $\sim$15 km s$^{-1}$ after Hanning smoothing. The data were calibrated and reduced following standard procedures in the Common Astronomy Software Applications (CASA, \citealt{McMullin2007}) package. 
We created a continuum image of each field with task $clean$ and the continuum was removed in the uv-plane based on the clean components using task $uvsub$. 
The \HI\ cubes were created with the same task ($clean$) using natural weighting to maximise the signal-to-noise ratio. GMRT has a $\sim4$ arcsec spatial resolution at 1.4 GHz. The data were tapered to lower resolution, a common $\sim15$ arcsec beam size, in order to improve the \HI\ brightness sensitivity. The $smallscale$ bias parameter was set to zero to avoid scaling of the \HI\ flux. A second continuum subtration was performed in the image plane with task $imcontsub$ using a second-order polynomial. This step was necessary to subtract residual continuum emission coming from very bright sources. The resulting GMRT spectra are shown in Fig. \ref{fig:Spectra} along with the GASS and ALFALFA spectra of the galaxies for comparison. The GMRT observations successfully recover the flux of the single dish measurements. After Hanning smoothing, the average 3$\sigma$ rms noise is 0.3 mJy beam$^{-1}$ channel$^{-1}$ in AGC 263047, and 0.2 mJy beam$^{-1}$ channel$^{-1}$ in the other three \HI\ cubes. We note that a noise enhancement can be observed in the \HI\ position-velocity diagrams in Fig. \ref{fig:HI_pv} as a result of the polynomial-based continuum subtraction in the excluded channels, corresponding to the velocity range covered by the \HI\ emission. 
The integrated intensity (zeroth-moment) and velocity field (first-moment) maps presented in Fig. \ref{fig:Collection} were created from the data cubes using task {\it immoments}. The zeroth-moment map was converted into column density following \cite{Morganti2007}.

\begin{figure*}
\begin{picture}(500,220)(0,0)
\put(30,0){\includegraphics[width=.434\textwidth]{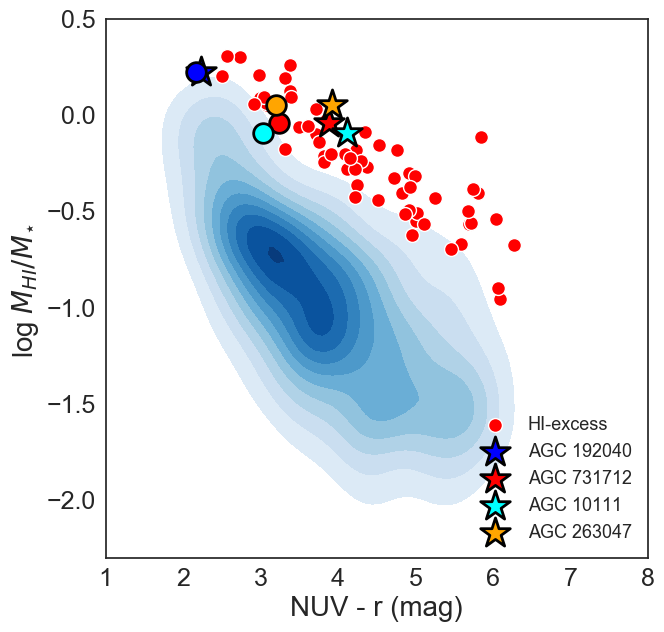}}
\put(260,-2){\includegraphics[width=.45\textwidth]{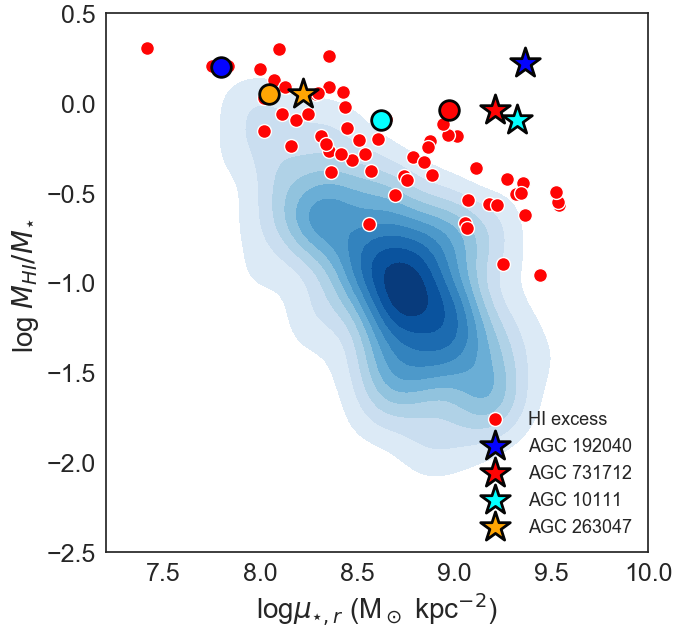}}
\end{picture}
\caption{Left panel: \HI\ gas fraction ($M_{\rm HI}/M_{\star}$) vs. NUV$-r$ colour relation of \HI-excess galaxies (red points). The blue shaded region marks the GASS sample. As in Fig. \ref{fig:GFplane}, stars and coloured circles (revised measurements) indicate the four \HI-excess galaxies discussed in this paper.
Right panel: \HI\ gas fraction vs. stellar mass surface density ($\mu_{\star,r}$) for the same samples.}
\label{fig:diagnostics}
\end{figure*}

\subsection{Optical spectroscopy}

The four galaxies were observed with the Echelette Spectrograph and Imager (ESI) on the 10m Keck II telescope at Mauna Kea on 29 Mar 2016. We used ESI in low-dispersion mode with a slit 1" wide and 8' long, aligned with the major axis of each galaxy. The average spectral resolution was R$\sim$3000 across our range from $\sim$4000A to $\sim$8000A. The spatial resolution along the slit direction is $\sim$1". We integrated for 6x1200s on each target. 

These longslit spectra were reduced with the standard processing tools in IRAF\footnote{IRAF is distributed by the National Optical Astronomy Observatory, which is operated by the Association of Universities for Research in Astronomy (AURA) under a cooperative agreement with the National Science Foundation.}, and the observations were corrected for overscan, bias, flat field, cosmic rays, and rotation. Multiple exposures of each target were averaged together and re-projected to a common wavelength scale using observations of calibration lamp spectra. 

For each target, we fit Gaussian profiles to the nebular emission and stellar absorption lines at each position along the slit and determine the barycentric velocity of each feature. All four galaxies in this sample show emission lines beyond their bright central regions, supporting the presence of star-forming regions as suggested by the UV images (see Fig. \ref{fig:Collection}). The ionized gas and stellar kinematics are compared to the \HI\ kinematics in Fig. \ref{fig:HI_pv}. 
The \HI\ position-velocity diagrams were extracted along the position angle and central coordinates of the Keck slit observations to perform this comparison.

\subsection{UV imaging}

We use GALEX FUV images to study the SFR across our sample and the NUV images to re-measure the colour of four galaxies (see Sec. \ref{sec:color_mustar}). The images of the four \HI-excess galaxies are retrieved through the Multimission Archive at Space Telescope Science Institute (MAST) system.
The Guest Investigator Database (GII) contains 1595s exposure observations both in FUV and NUV for AGC 192040.
The UV images of the other three \HI-excess galaxies are retrieved from the GALEX All Sky Survey (AIS). The AIS observed over 26,000 square degrees of sky with 100s exposure time, reaching a depth of m$_{AB}$ = 20.5. 
The FUV maps are shown in Fig. \ref{fig:Collection} (second column), overlayed with the \HI\ contours.


\vspace{-0.3cm}

\section{Analysis and Results}\label{Sec:Results}

\subsection{Revised colour and stellar surface density measurements}\label{sec:color_mustar}

The presence of blue outskirts in \HI-excess galaxies (Fig. \ref{fig:Collection}, left) warrant that colour gradients are involved in the modelMag photometry in SDSS\footnote{http://www.sdss3.org/dr8/algorithms/magnitudes.php}. As a result, the measured colour will strongly depend on the used aperture. 
Given that the SDSS measurements are optimised for the central regions where the galaxies are redder, here we re-measure the total NUV$-r$ colour in the four \HI-excess galaxies using large enough apertures that include the total flux. Given that overestimated (redder) NUV$-r$ colours result in a larger apparent disconnection between \HI\ and star-forming properties, we also test whether the \HI-excess in the gas fraction plane is not due to an aperture effect. 

We defined elliptical apertures manually to include the entire extent and having the same ellipticity and position angle of the outer UV disk of our targets. We calculate the total magnitude inside these NUV regions both for the NUV\footnote{https://asd.gsfc.nasa.gov/archive/galex/FAQ/counts$\_$background.html} and the SDSS\footnote{http://www.sdss3.org/dr8/algorithms/magnitudes.php} $r$-band images. The colours are corrected for Galactic extinction based on \cite{Wyder2007}. 
The newly measured NUV$-r$ colors are shown as function of gas fraction in Fig. \ref{fig:diagnostics} (left). The previous (catalog-based) measurements of the four galaxies are marked as stars, and the new, total NUV$-r$ measurements are shown as coloured circles. 
The GASS sample is represented by blue shaded regions in these plots, and \HI-excess galaxies are overplotted in red. 
Fig \ref{fig:diagnostics} clearly shows that \HI-excess galaxies occupy a different region than GASS in the log$M_{\rm HI}/M_{\star}$ vs. NUV$-r$ plane. 
The four \HI-excess galaxies analysed in this paper move bluewards after we include their outskirts.
However, the galaxies remain at the upper envelope of the colour vs. gas fraction relation. We also show the new location of the galaxies in the gas fraction plane, marked as coloured circles in Fig. \ref{fig:GFplane}. The galaxies move closer to the plane, however the \HI-excess is still $>2.5\sigma$ even after we include the contribution of the bluer disks.

Exclusion of outer disks can also contribute to the overestimation of the stellar surface density in galaxies. The SDSS Petrosian $r_{50,z}$ radii in the $z$ filter (used to compute $\mu_{\star}$ values in Fig. \ref{fig:GFplane}) range between 1-5 arcsec in the four galaxies, however the size of the low surface brightness disks are 30-50 arcsec in the SDSS $r$-band. In fact, the $z$-band filter is not sensitive enough to detect these faint outskirts at its 20.5 mag limit.
The highest signal-to-noise ratio in SDSS is measured in the $r$-band, providing images that reach $\sim$2 magnitudes deeper than the $z$-filter.
In the following we use the more sensitive SDSS $r$-band images to measure the $r_{{\rm eff}, r}$ (the radius at which half of the total light is emitted), and re-calculate the $\mu_{\star,r}$ in the selected sample. 
We note that even though deep imaging is available for three out of four galaxies, these are taken in different bands and there is no comparison sample available.

The $r_{{\rm eff}, r}$ is measured in the four galaxies based on the curve of growth technique described in \cite{Cortese2012}. In order to be consistent with the definition of Petrosian radii used by SDSS, surface brightness photometry was perfomed using circular apertures instead of following the ellipticity of each individual object. 
For consistency, the stellar mass density is re-calculated for the entire sample using catalog-based $r$-band measurements, such as $\mu_{\star,r}$ = $M_\star$ /(2$\pi$ $r_{50,r}^2$), where $r_{50,r}$ is the radius containing 50$\%$ of the Petrosian flux in the SDSS $r$-band in DR7.
The $\mu_{\star, r}$ measurements are shown as function of gas fraction in Fig. \ref{fig:diagnostics} (right).
The $\mu_{\star,r}$ values of the four galaxies based on the SDSS Petrosian $r_{50,r}$ are shown as stars, and we also show the new measurements based on $r_{{\rm eff}, r}$ marked as coloured circles. GASS and \HI-excess galaxies are marked as blue shaded regions and red points, respectively. We note that the difference in SDSS $r$-band and $z$-band Petrosian radius is less than 1 arcsec in about 90$\%$ of the sample, thus this plot shows the same trends when using the $z$-band radii.
Three out of the four galaxies analysed in this paper are clear outliers (star symbols) in Fig. \ref{fig:diagnostics} (right), however they move closer to the GASS distribution after the outskirts are accounted for (coloured circles). The $r_{{\rm eff}, r}$ shows a dramatic change in the case of AGC 192040, from a Petrosian $r_{50,z}$ = 1.45 arcsec to $r_{{\rm eff}, r}$ = 11 arcsec (with an error of 3 arcsec). The $r_{{\rm eff}, r}$ radius of the other three galaxies is up to 2.5 times larger than their Petrosian $r_{50,z}$. The galaxies are less extreme based on the revised $r$-band measurements, but they are still located at the upper envelope of the $\mu_{\star,r}$ vs. gas fraction relation. It is clear that the four \HI-excess galaxies are among the most gas-rich galaxies for their structure even after we include their outskirts.

For AGC 192040 we also see a clear segregation of the bulge and an extended disk component in the $r$-band image, and find that the SDSS magnitude measurements are mainly confined to its bulge. This suggest that the total stellar mass of the galaxy is underestimated. 
We measure the total flux of the galaxy and estimate the stellar mass of the disk based on the difference between the total and bulge magnitude. The outer disk represents only $\sim5\%$ of the bulge mass, showing that most of the total stellar mass is contained in the centre. We note that in Fig. \ref{fig:diagnostics} (right, blue circle) and Fig. \ref{fig:logGF_Mstar} (blue star) the revised, total stellar mass is shown for AGC 192040. For the other galaxies the bulges and disks are not as well defined, thus bulge-disk decomposition would be needed to measure the different components independently. 

When we compare the two panels in Fig. \ref{fig:diagnostics}, \HI-excess galaxies (red points) are stronger outliers with respect to the GASS sample in the NUV$-r$ vs. gas fraction relation than in the $\mu_{\star,r}$ one.
This suggests that colour is the main reason behind the \HI-excess selection in the gas fraction plane, but galaxy structure plays a secondary role.

\begin{figure}
\includegraphics[width=.43\textwidth]{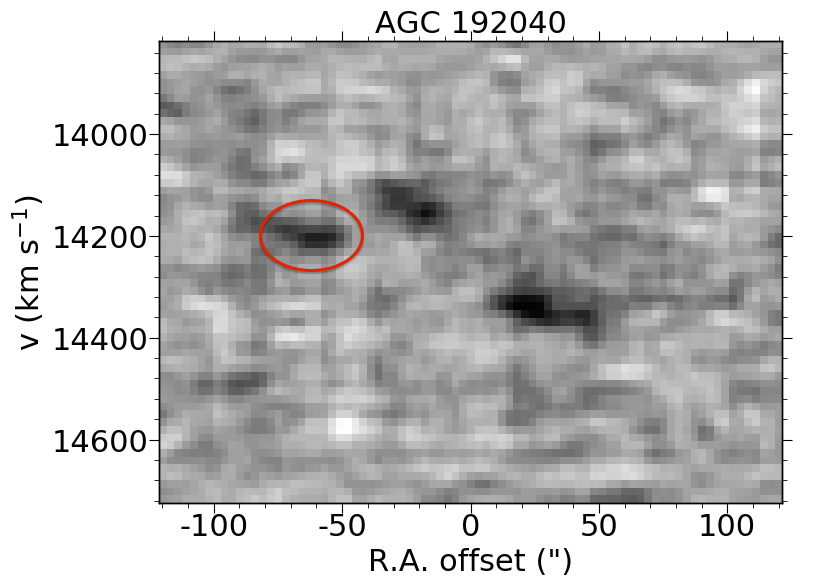}
\begin{center}\caption{\HI\ position-velocity diagram of AGC 192040 extracted along PA = 78.4 deg.
 The red ellipse marks the location of the \HI\ warp.} \label{fig:HI_pv_cloud}
\end{center}
\end{figure}

\subsection{Signatures of accretion in the multiwavelength data}\label{sec:multiwavelength}
 
In this Section we take advantage of our multiwavelength observations to look for signatures of gas accretion as a possible source of the \HI\ excess in our sample.

Figures \ref{fig:Collection} and \ref{fig:Collection_Optical} give an overview of the optical, UV and \HI\ properties of the sample. Deep optical imaging in Fig. \ref{fig:Collection_Optical} shows that a common feature of our \HI-excess galaxies is $\gtrsim$24 mag arcsec$^{-2}$ surface brightness stellar regions extending to large radii (up to 70 kpc radius in AGC 192040).
These are typical surface brightnesses for the diffuse light measured at the outskirts of galaxies where tidal streams and shells are often created as a result of galaxy interactions and mergers \citep{Janowiecki2010, Martinez2010, Duc2015}. The faint regions in Fig. \ref{fig:Collection_Optical} do not reveal shells or obvious stellar streams, instead they are distributed in spiral arms. The FUV images in Fig. \ref{fig:Collection} (second column) support the presence of star-forming regions, where star formation is embedded in \HI\ gas as indicated by the contours (in black).

In the \HI\ cube of AGC 192040 we identify a warp detached from the \HI\ disk (see red ellipse in the position-velocity diagram in Fig. \ref{fig:HI_pv_cloud}). The warped feature constitutes $\sim15\%$ to the total \HI\ mass in this galaxy, where stellar and UV components are both detected coinciding with the gas. 
Numerical simulations show that warps can be created when gas is accreted onto the disk with a slewed angular momentum in galaxies like the Milky Way \citep{Ostriker1989, Jiang1999}. However, tidal interactions can also cause disks to warp, and thus this feature is not a direct indication of gas accretion in AGC 192040. In Sec. \ref{Sec:env} we study the environement of our \HI-excess galaxies, allowing us to explore the latter scenario.

Fig. \ref{fig:Collection} (right panel) presents the velocity field and Fig. \ref{fig:HI_pv} the position-velocity measurements of the \HI\ gas. In Fig. \ref{fig:HI_pv} we also show the Keck measurements, where nebular emission lines arise due to the presence of ionized gas (cyan points) and absorption lines are coming from stellar photospheres (red points).
The agreement between the ionised and \HI\ gas kinematics suggests that these two gas phases are part of the same structure. Similar conclusions have been reached regarding the two gas phases in previous work by \cite{Morganti2006, Gereb2013, Serra2014}. The bulk of the \HI\ gas is in a rotating configuration in all galaxies. However, the position-velocity diagram of AGC 10111 suggests that the stars in the central bulge are counter-rotating with respect to the ionised/\HI\ disk. We measure the counter-rotation to be statistically significant by comparing the amplitude of the slope to the uncertainties from the Gaussian fits to absorption lines. Counter rotation is a strong indication that AGC 10111 has externally accreted its gas.

In summary, AGC 10111 is the only galaxy in our sample showing clear signatures of gas accretion in form of counter-rotation. No obvious evidence for gas accretion is found in the other galaxies.

\subsection{Star-forming properties of the \HI-excess sample}\label{Sec:UVprop}

In Paper I we showed that the star-forming processes in GASS 3505 are very inefficient, with gas depletion timescales of $\sim10^{11}$ yr. Such long timescales are typical for galaxy outskirts in general, where the column density of \HI\ is low \citep{Bigiel2010, Yildiz2015, Yildiz2017}. In AGC 192040, \cite{Lee2014} carried out molecular gas observations and detected $^{12}$CO (${\rm J}=1 \rightarrow 0$) observations to estimate molecular hydrogen gas masses and obtained log$M_{\rm H2}$ = 9.37 \msun, which translates into a molecular gas fraction of log$M_{\rm H2} /M_{HI}$ = -1.39. This number is very low compared to the COLD GASS distribution \citep{Saintonge2011}, where the average molecular gas fraction for a massive galaxy like AGC 192040 is log$M_{\rm H2} /M_{HI}$ = -0.5, suggesting that molecular gas formation is a very inefficient process in this galaxy. The other three galaxies in our sample do not have $M_{\rm H2}$ mass estimates available.

\begin{figure}
\includegraphics[width=.45\textwidth]{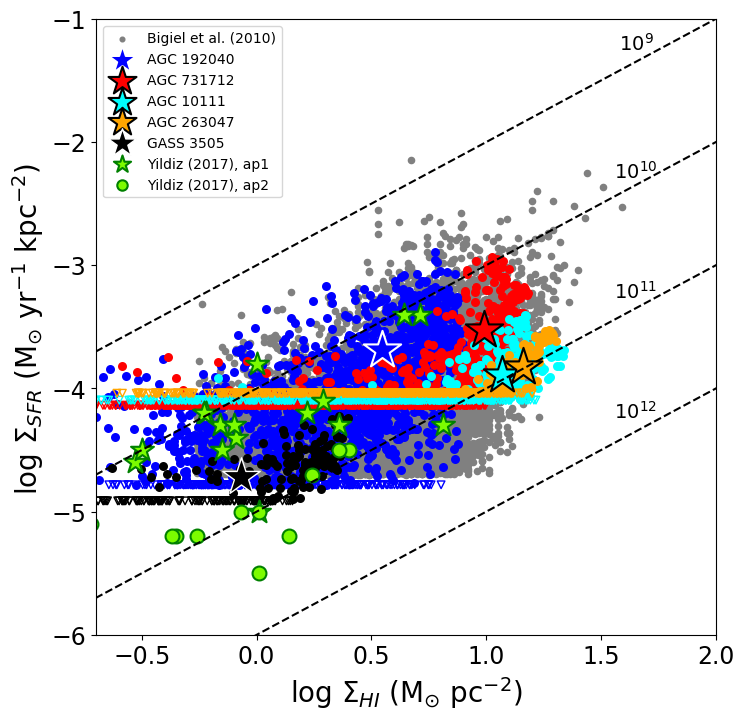}
\begin{center}\caption{SFR surface density vs. \HI\ surface density relation at the low end of the Kennicutt-Schmidt law. Star-forming regions in \HI-excess galaxies (where the FUV emission is above 3$\sigma$) are shown as colored dots, and triangles mark the 3$\sigma$ FUV upper limits. Star symbols mark the global measurements. The Bigiel et al. (2010) sample is shown in grey points, whereas the Yildiz et al. (2017) measurements are indicated by light green symbols. The black dashed lines mark constant gas depletion timescales of 10$^9 - 10^{12}$ yr.}\label{fig:KS}
\end{center}
\end{figure}

FUV emission is present in all galaxies in Fig. \ref{fig:Collection} (second column), coinciding with the highest column density \HI\ regions. We measure the SFR vs. \HI\ surface density relation (also known as the Kennicutt-Schmidt relation, \citealt{Schmidt1959, Kennicutt1989, Kennicutt1998}) in each pixel of the maps to learn about the efficiency of star formation in the \HI-excess sample.  
The FUV maps are background subtracted in IRAF, the photon counts are converted into flux, 
then into SFR, following \cite{Bigiel2010}. We smooth the maps to 15 arcsec to match the resolution of the \HI\ data. The pixel size is set to be 3 arcsec in both the \HI\ and FUV maps, corresponding to 1.5-2.7 kpc in the redshift range of the observations. 
Star-forming regions (where the FUV flux is higher than $3\sigma$) are marked as filled circles, upper limits (where the FUV flux is lower than $3\sigma$) as triangles in Fig. \ref{fig:KS}. Star symbols mark the global measurements estimated within the FUV radius. The corresponding SFR-s are reported in Table \ref{table:summary}. 

We show the results for our prototype \HI-excess galaxy, GASS 3505, along with the new measurements.
The sensitivity limit difference between the $\Sigma_{SFR}$ of AGC 192040 and GASS 3505 compared to the rest of the sample is due to their selection from deeper surveys. 
We also show the measurements by \cite{Bigiel2010} for the outskirts of spiral and dwarf galaxies, marked as grey circles. \cite{Yildiz2017} carried out a similar study using two different apertures that focus on the inner (1-3 $R_{\rm eff}$, aperture 1) and outer (3-10 $R_{\rm eff}$, aperture 2) parts of \atlas\ galaxies \citep{Serra2012}. These points are shown as light green stars and circles for the inner and outer apertures, respectively.  
\HI-excess galaxies follow the underlying distribution of previously studied samples, revealing inefficient star formation at their outskirts. The gas depletion timescales are between $10^{10} -10^{11}$ year as indicated by the dashed lines.

\begin{table*}
  \caption{Properties of galaxies in the environment of the \HI-excess sample. Col.1 Target AGC galaxy; Col.2 SDSS identification number; Col.3 SDSS redshift; Col.4; line-of-sight velocity difference relative to the target galaxy; Col.5 projected distance from the \HI-excess galaxy; Col.6 stellar mass; Col.7 \HI\ mass; Col.8 Membership to the target galaxy, where $1$ marks gravitationally bounded cases and $0$ marks non-members (see text). }\label{table:appendix}
 
   \begin{center}
    \begin{threeparttable}
     \begin{tabular}{l l l l l l l c} 		
\hline									     
Target AGC galaxy & SDSS ID  & z   & $v_{\rm los}$ & $d_{\rm proj}$       & log($M_{\star}$) & log($M_{\rm \HI}$) & Membership\\   
              &           &             &    (km s$^{-1}$)  &  (kpc)    &(\msun)          &  (\msun)          & \\
\hline	
AGC 192040 & SDSSJ094645.18+105414.0 & 0.0471            & 99$^{\dagger}$ & 818 &  10.13 & $<$9.4 & 0 \\ 
           & SDSSJ094706.15+104957.0 & 0.0474            & 24$^{\dagger}$ & 451 &  9.58  & $<$9.6 & 1 \\
AGC 731712 & SDSSJ112633.66+240100.3 & 0.0265 $^{\star}$ & 202            & 42  &  8.76  & 8.21   & 0  \\
           & SDSSJ112650.36+240613.2 & 0.0197 $^{\star}$ & 1832           & 167 &  9.07  & 8.87   & 0  \\
AGC 10111 & SDSSJ155836.37+130503.9 & 0.0344             & 55             & 314 &  10.01 & 9.69   & 1 

\end{tabular}
    \begin{tablenotes}
     \item[$\star$]  based on \HI\ velocity.
     \item[$\dagger$]  based on SDSS redshift.
    	\end{tablenotes}
    	   \end{threeparttable}
\end{center}
\end{table*}

\subsection{The environment of HI-excess galaxies}\label{Sec:env}

It has been known for several decades that high density environments, such as clusters, play a very important role when it comes to gas removal processes \citep{Giovanelli1983, Solanes2001, Cortese2011, Denes2016}. Low-density environments, on the other hand, provide favourable conditions for galaxies to accumulate gas \citep{Wang2015, Janowiecki2017}. Now we examine the observed volume around the four \HI-excess galaxies both in \HI\ and in the SDSS spectroscopic catalog. Our sample is selected against \HI\ confusion within the Arecibo beam, therefore we avoid the densest environments by selection.

\begin{figure*}
\begin{center}
\includegraphics[width=.55\textwidth]{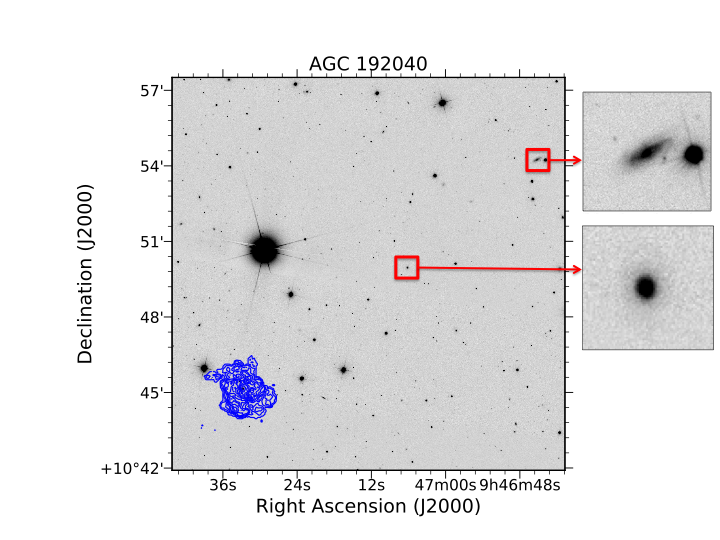}
\includegraphics[width=.55\textwidth]{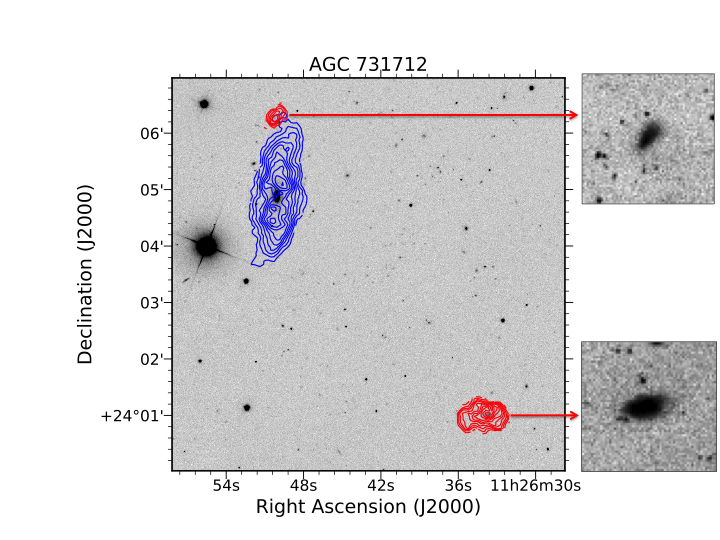} 
\includegraphics[width=.55\textwidth]{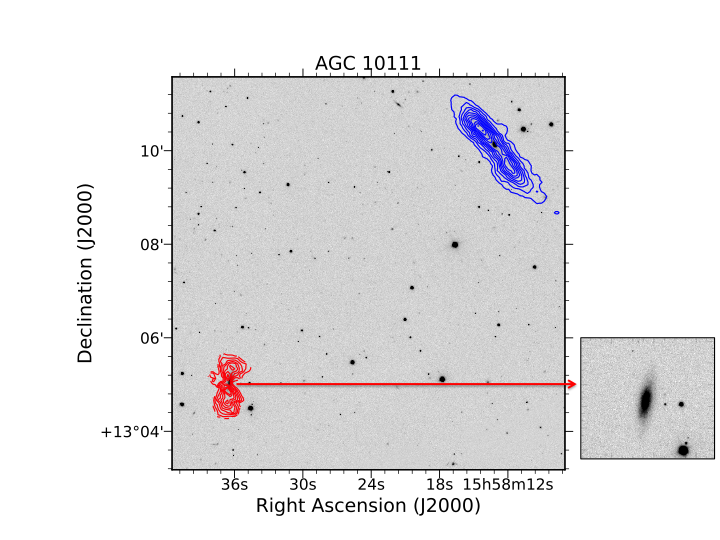}
\caption{The environment of \HI-excess galaxies AGC 192040, AGC 731712, and AGC 10111 from the top to the bottom. These galaxies are marked in blue in all cases, with the same contours as in Fig. \ref{fig:Collection}. $Top$: The environment of AGC 192040. The surrounding galaxies with \HI\ non-detections are marked as red boxes.
$Middle$: The environment of AGC 731712. The dwarf galaxies in the surrounding volume are shown in red. \HI\ contours associated with the dwarf in the north-eastern side of the galaxy range between (0.3 - 2)$\times$10$^{20}$ cm$^{-2}$ with steps of 0.3$\times$10$^{20}$ cm$^{-2}$, whereas the contours of the south-western galaxy are set to be between (0.3 - 6.6)$\times$10$^{20}$ cm$^{-2}$ with steps of 0.6$\times$10$^{20}$ cm$^{-2}$.
$Bottom$: An \HI\ detection in the environment of AGC 10111, marked in red. The red contours range between (0.35 - 9.5)$\times$10$^{20}$ cm$^{-2}$ with steps of 1.05$\times$10$^{20}$ cm$^{-2}$.}\label{fig:All_env}
\end{center}
\end{figure*}

The optical and \HI\ maps of the environment are shown in Fig. \ref{fig:All_env}. 
All galaxies have a few neighbours except for AGC 263047, thus we do not show a map of this galaxy in the Appendix. 
There are two galaxies in the vicinity of AGC 192040 (marked as red boxes in Fig. \ref{fig:All_env}, top panel) with available SDSS spectroscopy, however these are not detected in the GMRT \HI\ observations at the $3\sigma$ upper limit of log$M_{\rm HI} < 9.6$ \msun\ and log$M_{\rm HI} < 9.4$ \msun\ (assuming a line width of 200 \kms). 
The \HI\ non-detections are likely due to the fact that these galaxies are located in high-noise regions of the cube, far away from the phase centre where the sensitivity declines due to the effect of the primary beam.
In the environment of AGC 731712 we find two \HI\ detections with faint optical counterparts in SDSS. These dwarf galaxies do not have optical spectroscopy, similarly to the cases found in the neighbourhood of GASS 3505. We also find an \HI\ detection near AGC 10111 with available SDSS spectroscopy. The SDSS images of the environment are presented in Fig. \ref{fig:All_env}, where other galaxies detected in the \HI\ data cubes are shown with red contours, and non-detected ones are indicated by red boxes. The properties of these galaxies are summarized in Table \ref{table:appendix}. 

In the following analysis we explore whether the neighbours are gravitationally bound to the \HI-excess galaxies. A system is considered gravitationally bound if its potential energy exceeds the total kinetic energy, i.e., if $(G \times M)/r_{\rm sep} > 1/2 \times v^{2}$, where $G$ is the gravitational constant, $M$ is the total mass of the system, $r_{\rm sep}$ is the physical separation between the pair of galaxies, and $v$ is the velocity difference between these two \citep{Evslin2014}. The three-dimensional velocity can be approximated as $v \simeq \sqrt{3} v_{\rm los}$ for an isotropic system, where $v_{\rm los}$ is the line-of-sight velocity difference. The parameters used in these calculations are presented in Table \ref{table:appendix}. The total mass of every pair is calculated as the sum of $M_{\rm halo} + M_{\star} + M_{\rm HI}$, where $M_{\rm halo}$ is estimated using the stellar mass based on a study by \cite{Behroozi2013}.
In the last column of Table \ref{table:summary} we present the outcome of this analysis marking gravitationally bound galaxies with $1$, and non-members with $0$. We find that two of the neighbouring galaxies are gravitationally bound, however three other cases are not associated to \HI-excess galaxies. 
These examples support the results found for GASS 3505, that \HI-excess galaxies are typically found in low-density environments, i.e., in small groups or in isolation. Such environments are ideal for galaxies to build and preserve \HI-excess gas reservoirs.

\vspace{-0.3cm}

\section{Discussion}\label{Sec:Discussion}

In this Section we focus on discussing the $>2.5\sigma$ \HI\ excess in relation to the star-forming and structural properties in our sample.

The $\mu_{\star,r}$ vs. gas fraction relation in Fig. \ref{fig:diagnostics} (right panel) reveals that the \HI-excess sample (coloured circles) has high gas fractions for galaxies of the same structure. The four \HI-excess galaxies discussed in this paper have massive bulges (see optical images in Fig. \ref{fig:Collection}). The \HI\ content of bulge-dominated galaxies has been extensively studied in the last decades by numerous groups \citep{ Knapp1985, Wardle1986, vanDriel1991, vanGorkom1997, Sadler2002, Morganti2006, diSeregho2007, Oosterloo2007, Oosterloo2010, Grossi2009, Serra2012, Serra2014}. We note that these samples were selected based on morphological type as opposed to our gas fraction plane selection presented in this study. Among these \atlas\ is representative for the early-type galaxy population \citep{Serra2012}, and we use this sample to put the gas-richness of \HI-excess galaxies into context. \atlas\ has surveyed the \HI\ content of 166 early-type galaxies in- and outside of the Virgo cluster, and detected $10^7 - 10^{10}$ \msun\ of \HI\ in about half of the sample outside Virgo. The largest gas fraction among the settled disks of \atlas\ galaxies is log$M_{\rm HI}$/$M_{\star} = -0.8$ in the stellar mass range of GASS, whereas the gas fraction of the four galaxies studied in this paper varies between $-0.093 \leq$ log$M_{\rm HI}$/$M_{\star}$ $\leq 0.222$. Thus, \HI-excess galaxies are on average 10 times more \HI-rich compared to the \atlas\ sample.
This explains the outstanding position of the galaxies in Fig. \ref{fig:diagnostics} (right panel), showing that \HI-excess galaxies are the most gas-rich amongst bulge-dominated systems. So how can these galaxies keep such large gas reservoirs?

\begin{figure}
\includegraphics[width=.43\textwidth]{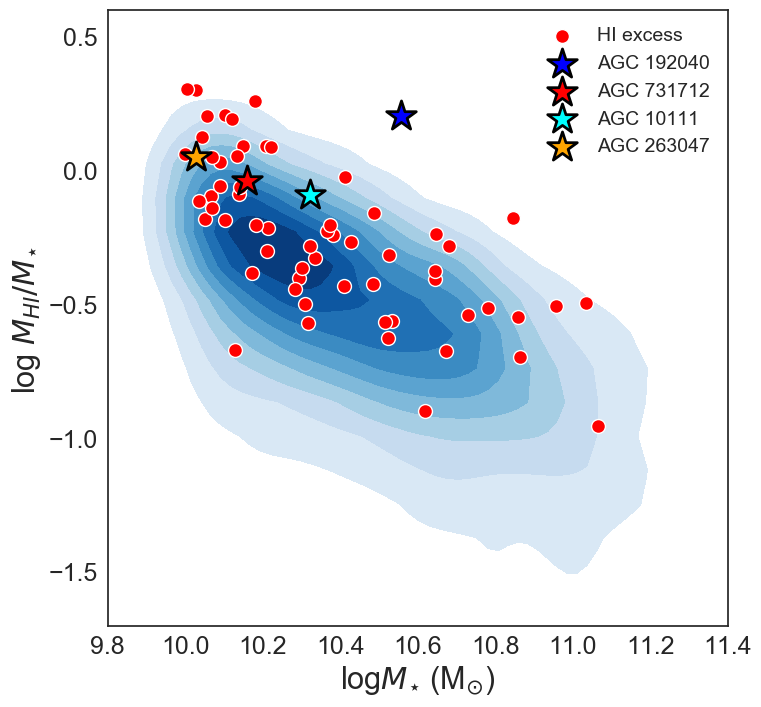}
\begin{center}\caption{Gas fraction ($M_{\rm HI}/M_{\star}$) vs. stellar mass ($M_{\star}$) relation of \HI-excess galaxies (red points and coloured stars), and the underlying GASS sample (blue shaded region). The $M_{\star}$ of AGC 192040 is based on the updated measurements.} \label{fig:logGF_Mstar}
\end{center}
\end{figure}

\HI-excess galaxies are distributed at the upper envelope of the NUV$-r$ vs. gas fraction relation in Fig. \ref{fig:diagnostics} (left panel), revealing a clear disconnection between \HI\ and star-forming properties. We explore whether this is due to the galaxies having too much gas, or due to inefficient star formation in the low surface brightness disks.
In Fig. \ref{fig:logGF_Mstar}, AGC 192040 is the only galaxy among the studied sample with exceptionally high gas fraction for its stellar mass. In fact, the log$M_{\rm HI}$(\msun) = 10.76 \HI\ mass of this galaxy is almost two times higher than its $M_{\star}$. 
There exist previous studies of non-morphologically-selected \HI-rich galaxies like Bluedisk \citep{Wang2013}, HIghMass \citep{Huang2014}, \cite{Lemonias2014}. None of these samples contain such high gas fractions in the log$M_{\star}$(\msun) $\sim 10.5$ stellar mass regime, however a few similarly gas-rich systems were found in the HIX sample by \cite{Lutz2017}. Looking at Fig. \ref{fig:logGF_Mstar}, the other three \HI-excess galaxies have normal gas fractions for their stellar mass, suggesting that these are not particularly gas rich for their stellar content. This leaves us with the conclusion that low star-forming efficiency pushes \HI-excess galaxies to the upper envelope of the NUV$-r$ vs. gas fraction relation.

Low star-forming efficiency is also supported by the KS relation in Fig. \ref{fig:KS}, where we show that the \HI-based gas depletion timescales at the current rate of star formation are of the order of 10$^{10} - 10^{11}$ years. This is 10 - 100 times longer than the average \HI-depletion time in "normal" SF galaxies ($\sim$3$\times$10$^9$ yr, but with a large scatter; \citealt{Schiminovich2010, Gereb2015}). The low molecular gas fraction detected in AGC 192040 (see Sec. \ref{Sec:UVprop}) suggests that CO formation is a very inefficient process in the extended disk of this galaxy. Similar results were found for the \HI-excess galaxy GASS 3505, where CO observations by the COLD GASS survey yielded a log$M_{\rm H2}$(\msun) $<8.93$ upper limit, corresponding to a log$M_{\rm H2}/M_{\rm HI}$ $<-0.97$ molecular gas fraction. Molecular gas observations are not available for other galaxies in our sample. However, based on the available data we speculate that something is preventing the collapse of gas into molecular clumps in \HI-excess galaxies.

In \cite{Gereb2016} we explored the multiwavelength properties of our prototype \HI-excess galaxy, GASS 3505, and found that its low star-forming efficiency can be attributed to low gas column densities in an extended \HI\ ring ($\sim$50 kpc radius). Using numerical simulations we showed that a GASS 3505-type galaxy can be created in a merger between a very gas-rich dwarf and a bulge-dominated system. Our results also suggest that multiple events are needed to explain the large \HI\ mass ($10^{9.9}$ \msun) measured in this galaxy.
Here, in Sec. \ref{sec:multiwavelength} we looked for signatures of external gas accretion in the current sample and found evidence for counter-rotation between the gas and stars in AGC 10111. 
Kinematic misalignment between the \HI\ and stellar component of bulge-dominated galaxies has been identified in previous work by \cite{vanGorkom1987, Bertola1992, Bettoni2001, Serra2014}. In fact, \cite{Serra2014} shows that about half of all large \HI\ disks around early-type galaxies are misaligned in the form of polar or counter-rotating gas. In our multiwavelength dataset the counter-rotation in AGC 10111 is the most clear evidence for gas accretion, emphasising the importance of the combination of optical and \HI\ kinematic information for gas accretion studies.

AGC 192040 has an extended, warped \HI\ disk, however we find no clear indications of gas accretion in this galaxy. Our analysis and results presented in Sec \ref{Sec:env} and Table \ref{table:appendix} show that AGC 192040 is part of a small group with at least one other member, thus the warp is likely caused by tidal interactions within the group environment.
AGC 192040 has the most prominent disk among the sample in terms of its large log$M_{\rm HI}$/$M_{\star}$ $= 0.222$ gas fraction, and 65 kpc \HI\ radius (70" at z=0.047517). This galaxy is comparable to a Malin 1-type low-surface brightness galaxy in terms of its gas richness \citep{Pickering1997, Lelli2010, Boissier2016}. AGC 192040 is less extreme than Malin 1 (which has a disk size of 130 kpc radius and \HI\ content of log$M_{\rm HI}$($M_{\odot}$) = 10.82), but it could be a similar type of object at slightly lower redshift (Malin 1 is at z=0.08). 
Theoretical work shows that an elevated specific angular momentum is able to keep the gas in extended, low surface density configuration like in Malin 1, where the gas becomes stable against gravitational fragmentation and consequent star formation \citep{Mo1998, Boissier2000, Kravtsov2013}. It has been suggested that high angular momentum is responsible for the accumulation of large \HI\ reservoirs in galaxies that are selected to be \HI-rich for their stellar content \citep{Huang2014, Lemonias2014, Lutz2017}. AGC 192040 satisfies the selection criteria of these \HI-rich samples based on Fig. \ref{fig:logGF_Mstar}, strengthening the likelihood that high angular momentum is responsible for keeping the gas in an extended, \HI-rich configuration in AGC 192040.

In the remaining two galaxies (AGC 731712 and AGC 263047) the settled nature of the gas suggests that the \HI\ disks have been in place for at least one orbital time. This timescale is a few Gyr for massive galaxies ($>10^{10}$ M$_{\odot}$) with extended disks (tens of kpc-s) like the ones observed in our \HI-excess sample.
Despite the wealth of multi-wavelength data, it is still unclear how these galaxies gained and preserved their gas. AGC 731712 and AGC 263047 seem to be \atlas-type galaxies in terms of their large bulge-to-total ratios and extended \HI\ disks, however with larger \HI\ masses and gas fractions.
Numerical simulations are able to reproduce bulge-dominated galaxies with extended, gas-rich disks via mergers between two gas-rich galaxies \citep{Naab2001, Barnes2002, Springel2005, Robertson2006}. 
Mergers over a large range of mass ratios have been suggested to be responsible for the formation mechanisms for galaxies with an anomalous \HI\ content, like the ones observed in \atlas\ and other early-type galaxies \citep{Serra2006, Oosterloo2010, Serra2012}.

In summary, the five \HI-excess galaxies studied in this work and Paper I seem to be a mixed population, with two examples of external accretion (AGC 10111 and GASS 3505), and one galaxy showing signs of recent tidal interactions (AGC 192040), but it appears that the bulk of the \HI\ disks have been built at least a couple of Gyr ago. The diffuse nature of the extended outskirts, combined with low star-forming efficiencies points to a gradual build-up of the low surface brightness disks over an extended period of time like in a Malin 1-type galaxy \citep{Boissier2016}. \HI-excess galaxies are 1-2 orders of magnitude less efficient at converting their gas into stars than normal star-forming galaxies. However, \HI-excess galaxies are definitely not red and dead. There is usually a quick transition assumed to take place in galaxies as they move from the blue cloud to the red sequence in the color magnitude diagram \citep{Bell2004, Wyder2007, Schawinski2014}. Our sample reveals examples of galaxies at this intermediate stage. Based on the large gas masses and low star-forming efficiencies involved, the galaxies will remain in this state for a long time (more than 10 Gyr) as opposed to being transitional objects.

Thus, it appears that our selection based on the GASS gas fraction plane yields very unique and interesting objects. In order to understand the formation and evolution of these galaxies, hydrodynamic simulations of large volumes are needed in a cosmological context where the accretion processes in galaxies and merger histories can be tracked. Both the Evolution and Assembly of GaLaxies and their Environments (EAGLE, \citealt{Schaye2015}) and Illustris \citep{Nelson2015} simulations cover a volume of roughly 100 Mpc on a side. The volume density of \HI-excess galaxies is of the order of 10$^{-5}$ Mpc$^{-3}$ based on the sky regions covered by ALFALFA and GASS (in the redshift range $0.01 < z < 0.05$), thus the simulated volumes are clearly not enough to yield a statistically representative sample. 
Observationally, metallicity measurements can provide additional information on the build-up of the extended outksirts. Metal enrichment and the age of the stellar populations are strongly related through star formation, thus metallicity measurements can help to constrain the formation timescales involved in the build-up of \HI-excess galaxies \citep{Moran2010}.

\vspace{-0.3cm}

\section{Summary and Conclusions}\label{Sec:Summary}

This paper has focused on studying the deep optical, \HI, star-forming, and environmental properties of \HI-excess galaxies, a sample selected from gas fraction scaling relations based on the GASS survey. The ultimate goal of our study is to learn about the physical processes behind the build-up of \HI-excess galaxy disks.

\HI-excess galaxies are the most gas-rich among bulge-dominated systems, as reflected by their location in the $\mu_{\star,r}$ vs. gas fraction relation. We find that a few cases in our sample might represent scaled-up versions of \atlas-type galaxies in terms of their gas fraction.
\HI-excess galaxies are also very gas rich for their NUV$-r$ colours due to a disconnection between their large amounts of gas and low level of star formation.  

Based on the settled nature of the \HI\ distribution it appears that the bulk of the gas disks have been built a couple of Gyr ago. However, the galaxies are a mix in terms of the physical processes that keep their gas stable. AGC 192040 is a good candidate for being a Malin 1-type low surface brightness galaxy with its extended \HI\ disk (65 kpc) and large gas fraction (logM$_{HI}$/M$_{\star}$ = 0.222), where large specific angular momentum is preserving the gas in an extended configuration. Counter-rotation between the gas and stars in AGC 10111 is a clear indication that the gas in this galaxy is of external origin. 

What \HI-excess galaxies have in common is the contrast between their large gas fractions and low level of star formation, corresponding to gas depletion timescales of $10^{10} -10^{11}$ year. In fact, this sample represents a long-lived, intermediate stage between actively star-forming and bulge-dominated galaxies with no star formation. 
In conclusion, our detailed study of five representative \HI-excess galaxies (including GASS 3505) indicates an external origin for the \HI\ gas in two cases (GASS 3505 and AGC 10111), but it remains unclear how the other three systems obtained their large gas reservoirs. Numerical simulations of large cosmological volumes are needed to track the formation and evolution of these unique objects.


\vspace{-0.3cm}
\section{Acknowledgements}
We wish to thank the referee for constructive input on the manuscript. 
KG, SJ, BC and LC acknowledge support from the Australian Research Council's Discovery Projects funding scheme (DP130100664 and DP150101734).
BC is the recipient of an Australian Research Council Future Fellowship (FT120100660). 
The National Radio Astronomy Observatory is a facility of the National Science Foundation operated under cooperative agreement by Associated Universities, Inc.
We thank the staff of the GMRT that made these observations possible. GMRT is run by the National Centre for Radio Astrophysics of the Tata Institute of Fundamental Research.
This research has made use of the services of the ESO Science Archive Facility. Based on observations collected at the European Organisation for Astronomical Research in the Southern Hemisphere under ESO programme 088.B-0253(A).
(Some of) The data presented herein were obtained at the W.M. Keck Observatory, which is operated as a scientific partnership among the California Institute of Technology, the University of California and the National Aeronautics and Space Administration. The Observatory was made possible by the generous financial support of the W.M. Keck Foundation. The authors wish to recognize and acknowledge the very significant cultural role and reverence that the summit of Mauna Kea has always had within the indigenous Hawaiian community.  We are most fortunate to have the opportunity to conduct observations from this mountain.



\end{document}